\documentclass[pra,showpacs,preprintnumbers,tightenlines,superscriptaddress,twocolumn]{revtex4}

\usepackage{graphicx}% Include figure files
\usepackage{dcolumn}% Align table columns on decimal point
\usepackage{bm}% bold math
\usepackage{epsfig}
\usepackage{stmaryrd}
\usepackage{dsfont}
\usepackage{bbm}
\usepackage{amsmath}
\usepackage{amsfonts}
\usepackage{amssymb}
\usepackage{amscd}
\usepackage{amstext}

\newcommand{\expec}[1]{\langle #1 \rangle}

\newcommand{\casea}{(i)}
\newcommand{\caseb}{(ii)}
\newcommand{\casec}{(iii)}
\newcommand{\cased}{(iv)}

\newcommand{\tabvspace}{\Big.}

\newcommand{\sub}[2]{{#1}_{\mbox{\!\! \scriptsize #2}}}

\newcommand{\bv}[1]{\mathbf{ #1 }}

\def\beq{\begin{equation}}
\def\eeq{\end{equation}}
\def\nnl{\\[0.15cm] \nonumber}

\def\CR{\nonumber\\[0.15cm]}
\def \gap{\:\:\:\:}
\newcommand{\fref}[1]{Fig.~\ref{#1}}
\newcommand{\frefp}[2]{Fig.~\ref{#1}~(#2)}

\newcommand{\eref}[1]{Eq.~(\ref{#1})}

\newcommand{\sref}[1]{section~\ref{#1}}

\newcommand{\cref}[1]{chapter~\ref{#1}}

\newcommand{\Cref}[1]{Chapter~\ref{#1}}
\newcommand{\tref}[1]{table~\ref{#1}}

\newcommand{\bref}[1]{(\ref{#1})}

\begin{document}

\title{Correlations of Rydberg excitations in an ultra-cold gas after an echo sequence}
\author{S.~W\"uster}
\affiliation{Max Planck Institute for the Physics of Complex Systems, N\"othnitzer Strasse 38, 01187 Dresden, Germany}
\email{sew654@pks.mpg.de}
\author{J.~Stanojevic}
\affiliation{Max Planck Institute for the Physics of Complex Systems, N\"othnitzer Strasse 38, 01187 Dresden, Germany}
\author{C.~Ates}
\affiliation{Max Planck Institute for the Physics of Complex Systems, N\"othnitzer Strasse 38, 01187 Dresden, Germany}
\author{T.~Pohl}
\affiliation{Max Planck Institute for the Physics of Complex Systems, N\"othnitzer Strasse 38, 01187 Dresden, Germany}
\author{P.~Deuar}
\affiliation{Institute of Physics, Polish Academy of Sciences, Aleja Lotnik\'{o}w 32/46, 02-668 Warsaw, Poland}
%\affiliation{Universit{\'e} Paris-Sud, LPTMS, UMR8626, B{\^a}t.~100, 91405 Orsay cedex, France}
%\affiliation{CNRS, LPTMS, UMR8626, B{\^a}t.~100, 91405 Orsay cedex, France}
\author{J.~F.~Corney}
\affiliation{ARC Centre of Excellence for Quantum-Atom Optics, School of Mathematics and Physics, University of Queensland, Brisbane QLD 4072, Australia}
\author{J.~M.~Rost}
\affiliation{Max Planck Institute for the Physics of Complex Systems, N\"othnitzer Strasse 38, 01187 Dresden, Germany}

\begin{abstract}
We show that Rydberg states in an ultra-cold gas can be excited with strongly preferred nearest-neighbor distance if densities are well below saturation. The scheme makes use of an echo sequence in which the first half of a laser pulse excites Rydberg states while the second half returns atoms to the ground state, as in the experiment of Raitzsch {\it et al.} [Phys. Rev. Lett. {\bf 100} (2008) 013002]. Near to the end of the echo sequence, almost any remaining Rydberg atom is separated from its next-neighbor Rydberg atom by a distance slightly larger than the instantaneous blockade radius half-way through the pulse. These correlations lead to large deviations of the atom counting statistics from a Poissonian distribution. Our results are based on the exact quantum evolution of samples with small numbers of atoms. We finally demonstrate the utility of the $\omega$-expansion for the approximate description of correlation dynamics through an echo sequence.
\end{abstract}
\maketitle

%%%%%%%%%%%%%%%%%%%%%%%%%%%%%%%%%%%
\section{Introduction}

\begin{figure}[htb]
\centering
\epsfig{file={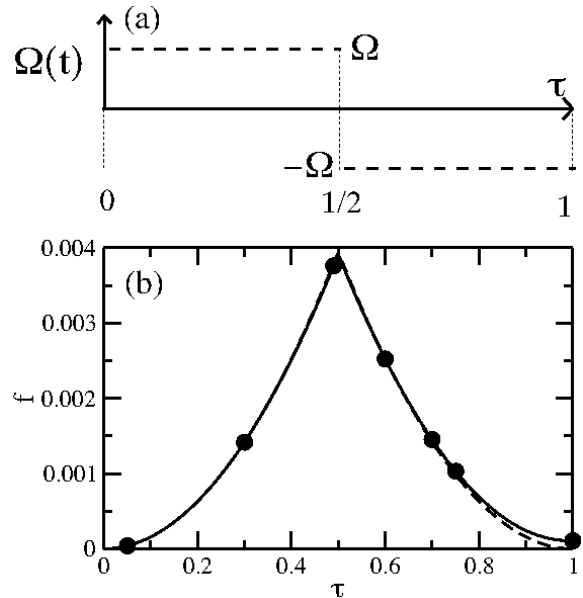},width=0.89\columnwidth} 
\caption{Illustration of an echo excitation sequence and evolution of the Rydberg fraction. For a sequence of duration $T$, we use the fractional time $\tau=t/T$. (a) Rabi frequency $\Omega w(t)$ [see \sref{ExcDyn}] during the sequence. (b) Fraction of excited state atoms, $f$, for a density $\rho_{0} V_{b0}=26.2$ (solid) from a numerical solution of Schr{\"o}dinger's equation. Here, $V_{b0}=4\pi r_{b0}^3 /3$ is a simple estimate of the saturated blockade volume. We also show the non-interacting result for the excited fraction (dashed). The symbols ($\bullet$) mark times for which we display correlations in \fref{Echo_tevol}.
\label{Echo_numbers}}
\end{figure}

When atoms within an ultra-cold gas are excited to Rydberg levels, they experience long-range interactions which can block further excitations, leading to a strongly correlated many-body state. This effect might be useful for quantum information \cite{lukin:quantuminfo} as well as for fundamental studies of many-body physics and was observed in several experiments \cite{tong:blockade,johnson:rabiflop,singer:blockade,heidemann:strongblockade,heidemann:rydberg,cubel:statistics,raitzsch:echo, younge:echo,gaetan:twoatomblock,urban:twoatomblock}. The coherence of the excitation process in a bulk gas has been experimentally demonstrated using an echo-technique \cite{raitzsch:echo,younge:echo}: After exciting atoms to Rydberg states, it was possible to de-excite them following a $\pi$-phase shift of the excitation laser. The basic scheme is illustrated in \fref{Echo_numbers}.
After the echo sequence, a certain fraction of the atoms remains in the excited state, owing to effects of the interaction. This has been modeled theoretically using the super-atom approach \cite{hernandez:echo}. 

Neither experiment nor theory has considered the dynamics of Rydberg-Rydberg correlations during such an echo-sequence. Here, we show that strong correlations of atoms separated by a characteristic distance $r_{0}$ are induced in its course. This distance $r_{0}$ is slightly larger than the instantaneous blockade radius at the moment when the laser phase is flipped. At the \emph{instantaneous blockade radius} $r_{b}(t)$ the Rydberg state density-density correlation function drops sharply to zero. Initially $r_{b}(t)$ equals zero, later growing towards its saturation value $r_{b0}\sim(C_{6}/\Omega)^{1/6}$, as the longer duration of the pulse allows an increasingly finer energy resolution.

After most Rydberg atoms that were excited in the first half of the pulse returned to the ground state, the majority of the remainder has their nearest excited neighbor at a distance between $r_{0}$ and $1.5 r_{0}$. The strength of this correlation signal is proportional to $\rho^{-2}$, where $\rho$ is the atomic density. 

Our results imply that the echo technique can be used to manipulate the nearest-neighbor distribution in the Rydberg fraction of the gas. Such manipulations could be used for example to initiate dynamics due to dipole-dipole forces \cite{cenap:motion} from a well specified non-equilibrium state.

Similar correlations between Rydberg atoms are created if a two-step excitation scheme with a strongly decaying intermediate state is used~\cite{cenap:antiblockade,cenap:manybody}. In that case they are due to the Autler-Townes splitting of the intermediate level.

The peculiar nature of the nearest neighbor distribution function of Rydberg atoms after an echo pulse could be seen experimentally in the kinetic energy spectrum of ions after field ionization of the remaining Rydberg fraction. Alternatively 
one could measure deviations of the atom counting statistics from a Poissonian distribution~\cite{cenap:mandelQ}. Both these suggestions are discussed further in \sref{signatures}.

%Also in the course of fully developed Rabi-Oscillations between ground and excited state components, we find a periodic preference of a certain nearest-neighbor distance. These oscillations of the correlation function accompany its relaxation to the equilibrium shape in the presence of laser-coupling, as described recently in \cite{weimer:critical}.  

Our results are based on solutions of the many-particle Schr{\"o}dinger equation, which we also use to benchmark the recently proposed $\omega$-expansion \cite{jovica:omegexp1,jovica:omegexp2}. Both methods are briefly described in \sref{Methods}. 
The ensuing correlation dynamics are presented in \sref{correldyn}. Possible ways to detect the pairing effect are discussed in \sref{signatures}. In \sref{interact_inhomog} we take a closer look at the effects of density variations on our results and in \sref{omegaexp} we directly compare many-body quantum simulations with the $\omega$-expansion. Finally, we conclude in \sref{conclusions}.

%%%%%%%%%%%%%%%%%%%%%%%%%%%%%%%%%%%
\section{Methods}
\label{Methods}

%%%%%%%%%%%%%%%%%%%%%%%%%%%%%%%%%%%
\subsection{Exact quantum dynamics}
\label{ExcDyn}

We consider a system of $N_{0}$ atoms at fixed locations $x_{i}$, described by the following Hamiltonian in atomic units:
\begin{align}
\hat{H}&= \frac{1}{2} \sum_{i=1}^{N_{0}}\Omega w(t) (\sigma_{eg}^{i} +\sigma_{ge}^{i})
\CR
&+ \Delta\sum_{i=1}^{N_{0}} \sigma_{ee}^{i} + \sum_{i,j;j>i}^{N_{0}} \kappa_{ij}\sigma_{ee}^{i} \sigma_{ee}^{j}.
\label{Hamiltonian}
\end{align}
Each atom can either be in its ground state $|g \rangle$ or an excited state $|e \rangle$, the Rydberg state. The atoms in $|e \rangle$ experience long range interactions $\kappa_{ij}=-C_{6}/|x_{i}-x_{j}|^6$. Laser induced transitions between the levels occur with Rabi-frequency $\Omega\in\mathbb{R}$, detuning $\Delta$ and the temporal profile of the pulse $w(t)$. The operators $\sigma_{ab}^{i}$ act as $\sigma_{ab}^{i}= |a \rangle \langle b|$ on the subspace of atom $i$ and as unity on others. We use atomic units unless otherwise indicated.

We expand the many-body wave function as $|\Psi(t)\rangle=\sum_{\bv n} c_{\bv n}|\bv n\rangle$, where $\bv n$ is a vector with $N_{0}$ elements $n_{j}\in \{0,1\}$ which describe whether atom $j$ is in the ground ($0$) or excited ($1$) state. Schr{\"o}dinger's equation (SE) then takes the form: 
\begin{align}
i \dot{c}_{\bv n}&=\left[ \Delta \sum_{i}^{N_{0}} n_{i}  +  \sum_{i,j; i> j}^{N_{0}} \kappa_{ij} n_{i} n_{j} \right]  c_{\bv n} 
\CR
&+ \frac{1}{2} \sum_{i}^{N_{0}} 
\left(    \Omega w(t) c_{\bv n, \downarrow i}    +  \Omega^* w^*(t) c_{\bv n, \uparrow i} \right),
\label{SE1}
\end{align}
where $c_{\bv n, \downarrow i}$ ($c_{\bv n, \uparrow i}$) is the coefficient of the state that is reached from $|\bv n\rangle$ by lowering (raising) the $i$'th atom if this is possible, and $0$ otherwise.

For the numerical solution we convert \eref{SE1} to a rotating frame for the variables
\begin{align}
\tilde{c}_{\bv n}&= e^{-i t \sum_{i,j;  i> j}^{N_{0}} \kappa_{ij}n_{i} n_{j} } c_{\bv n}.
\label{rotframe}
\end{align}
To render simulations of ensembles with hundreds of atoms possible, we remove all states whose interaction energy is larger than some cut-off $\sub{E}{cut}$ from the Hilbertspace (see eg.~\cite{weimer:critical,younge:beyond_pairwise}). We further limit the number of simultaneously excited atoms to $\sum n_{j}\leq \sub{N}{max}$. All numerical results are checked for convergence with respect to variations of $\sub{N}{max}$ and $\sub{E}{cut}$.

%%%%%%%%%%%%%%%%%%%%%%%%%%%%%%%%%%%
\subsection{The $\omega$-expansion}
\label{Omegaexp}

In the experiment \cite{raitzsch:echo}, the sign of the Rabi-coupling was flipped well before the excitation of Rydberg atoms reached the saturation limit imposed by the blockade effect. In such a situation one could attempt to solve the quantum dynamics of the gas by a series expansion in $\omega=\Omega t$ \cite{jovica:omegexp1}, where $\omega=0.25$ for the experimental conditions. 

From \eref{Hamiltonian}, we can derive equations of motion for the operators $\sigma_{ab}^{i}$. We expand these operators in a series  
 $\sigma_{ab}^{i}= \sum_{n} \omega^n \sigma_{ab}^{i(n)}$. It is possible to obtain recursive equations expressing $\sigma_{ab}^{i(n)}$ by $\sigma_{a'b'}^{i(m)}$ for $m<n$. In our case the lowest order approximations of the $\sigma_{ab}^{i}$ already offer interesting insight into correlation dynamics.   

In the following we are interested in an echo-type pulse. Assuming a pulse-length $T$, we rescale our variables 
\begin{align}
&\tau=t/T,\gap  \omega= \Omega T,  \gap  k_{ij} = \kappa_{ij}\ T, \gap  f(\tau)=w(t/T).
\label{rescaling}
\end{align}
The echo pulse shown in \fref{Echo_numbers} is then given by
$
f(\tau)= \Theta(1/2-\tau)-\Theta(\tau-1/2),
$
where $\Theta(t)$ is the Heaviside function. We further define $F(\tau)= \int_{0}^{\tau}d\tau' f(\tau')= \tau\Theta(1/2-\tau)+ [1-\tau]\Theta(\tau-1/2)$.

According to the $\omega$-expansion the leading- (LO) and next-to-leading- (NLO) order expressions for the excited state fraction $P_{e}(\tau)$ are given by:
\begin{align}
&\sub{P}{e}^{(LO)}(\tau)=\omega^2\langle \sigma_{ee}^{i(2)} \rangle=\omega^2 \frac{|F(\tau)|^2}{4},
%\CR=\frac{\omega^2}{4}  \tau^2
%\left[ \Theta(\half-\tau)+ (1-\tau)^2 \Theta(\tau-\half)\right],
\label{PeLO}
\\
&\sub{P}{e}^{(NLO)}(\tau)=\omega^2\langle \sigma_{ee}^{i(2)} \rangle + \omega^4 \langle \sigma_{ee}^{i(4)} \rangle
\CR
&\:\:\:\:\:\:\:\:\:\:\:\:\:\:\:\:\:\:\:\:\:=\sub{P}{e}^{(LO)} - \omega^4 (I_{41} + I_{42}),
\label{PeNLO}
\\
&I_{41}= \frac{|F(\tau)|^4}{16}-\mbox{Re}\left[
\frac{F(\tau)}{8}\int_{0}^{\tau} d\tau_{1} f(\tau_{1})^* F(\tau_{1})^2 
\right],
\\
&I_{42}=\frac{1}{4}\sum_{i\neq j} \mbox{Re}\bigg[
\int_{0}^{\tau} d\tau_{1} f(\tau_{1})[F(\tau) -2 F(\tau_{1})] 
\CR
& \gap\times \int_{0}^{\tau_{1}} d\tau_{2} f^*(\tau_{2})F^*(\tau_{2})\left(e^{i(\tau_{1} - \tau_{2})k_{ij}}-1 \right)
 \bigg].
\end{align}
The Rydberg-Rydberg correlation function is defined by 
\begin{align}
 g^{(2)}(i,j) &\equiv \frac{\langle \sigma_{ee}^{i}(\tau)\sigma_{ee}^{j} (\tau)\rangle}{\langle\sigma_{ee}^{i} (\tau)\rangle\langle\sigma_{ee}^{j}(\tau) \rangle},
\label{g2defn}
\end{align}
with leading order approximation in the $\omega$-expansion
\begin{align}
 g^{(2)}_{LO}(i,j) &= \frac{4\left| \int_{0}^\tau d\tau_{1} \:\: e^{i \tau_{1} k_{ij}} f(\tau_{1})F(\tau_{1})  \right|^2}{|F(\tau)|^4}.
\label{g2anal}
\end{align}
\begin{table}
\begin{center}
\begin{tabular}{|c|cccc|}
\cline{1-5}
\tabvspace
case      & {\casea} &  {\caseb} & {\casec} & {\cased}  \\
\cline{1-5}
$\tabvspace N_{0}$  & $20$ & $84$ & $125$ & $50$  \\
$L/r_{b0}$  & $2.5$ & $2.38$ & $2.38 $ & $2.38$  \\
$\rho_{g}(r)$     & $\rho_{0}$ & $\rho_{0}$ & $\rho_{0}$  & $\rho_{0}\exp{[-2r^2/\sigma^2]}$ \\
$\rho_{0} V_{b0}$     & $5.4$ & $26.2$ & $39$ & $26.2$ \\
$\rho_{0}/10^{10}[cm^{-3}]$     & $1.3$ & $6.3$ & $9.4$ & $6.3$ \\
\cline{1-5}
\end{tabular}
\end{center}
\caption{Parameters for the scenarios modeled in this article. In cases {\casea}-{\casec} the atoms are homogeneously distributed in a cubic box of side-length $L$ with periodic boundary conditions \cite{footnote:numerics2}. Case {\cased} is a gaussian cloud as shown, with $\sigma=1.6 r_{b0}$. This ensures the same peak-density as the homogenous density of case~{\caseb}. The relevant dimensionless measure of the density is $\rho_{0} V_{b0}$. For the parameters described in the text this would correspond to the values in the last row.
\label{scenarios}}
\end{table}

Calculating higher order corrections to these quantities, although possible in principle, is more tedious than justified. The expressions above can be explicitly evaluated for a homogeneous system with van-der-Waals interaction
%
%\begin{align}
$\kappa_{ij}=-C_{6}/|x_{i}-x_{j}|^6.
$
%\label{potential}
%\end{align}
%
The sum in $I_{42}$ is replaced by $\sum_{i\neq j} \rightarrow \int d^3x \rho(x)$, corresponding to an ensemble average. Throughout the paper we write $\rho$ for the full density profile and $\rho_{0}$ for the peak density. If we evaluate \eref{PeNLO} at $t=T$ for a homogeous system, we can obtain the echo signal, i.e.~the final Rydberg fraction
\begin{align}
f_{e}=P_{e}(T)= \frac{2\omega^4 \pi^{3/2}\sqrt{|C_{6}|} }{2835}\left(8\sqrt{2}-9\right)\rho_{0}.
\label{echosig}
\end{align}
Other analytical expressions that can be obtained are not illuminating and hence omitted here. However, we compare their predictions with direct numerical simulations in \sref{omegaexp}. A more detailed description of the $\omega$-expansion can be found in \cite{jovica:omegexp1}.

To estimate which order of the expansion in $\omega$ is required for a given scenario, we can consider the excited fraction of a fully blockaded sample of $N_{b}$ atoms assuming a square pulse,
\begin{align}
P_{e}(t)= \frac{1}{N_{b}}\sin^2[\sqrt{N_{b}} \Omega t/2]=\sum_{n} P^{(n)}_{e} (\Omega t)^n,
\label{peexp}
\end{align}
which is correctly reproduced by the expansion in the limit of infinite interactions \cite{jovica:omegexp1}. Roughly knowing the expected $N_{b}$ for a system and the maximal $\Omega t$, we can estimate how many terms of series expansion \eref{peexp} are required. This will be used in \sref{omegaexp}.

%%%%%%%%%%%%%%%%%%%%%%%%%%%%%%%%%%%
\section{Correlation dynamics}
\label{correldyn}

For $N_{0}$ atoms, homogenously distributed in a cube of volume $L^3$, we can see that the physics of our problem
is governed by two parameters, namely $\omega= \Omega T$ and $U=C_{6} T/ L^6$. Throughout this paper we usually employ $\omega=0.25$ and $U=-0.0014$. This corresponds, for example, to a $^{87}$Rb gas in which states with principal quantum number $\sub{n}{Ryd}=41$ are excited via a transition with Rabi-frequency $\Omega=2 \pi \times 0.1 \:\mbox{MHz}$ during a time of  $400$ ns. The side-length of the box would be $L=11\mu$m. For these parameters we then vary the number of atoms as listed in \tref{scenarios}, leading to densities of the order of $5\times 10^{10}$ cm$^{-3}$.  Compared to the experiment \cite{raitzsch:echo}, these parameters lead to a substantially weaker blockade with maximally about $N_{b}=40$ atoms per blockade-sphere. This facilitates both, the physics of interest here and the numerical simulation. We will use the saturated two-atom blockade radius $r_{b0}=(C_{6}/\Omega)^{1/6}$ as length scale and express densities using the simple estimate for the blockade volume $V_{b}=4\pi r_{b0}^3 /3$.

\begin{figure}
\centering
\epsfig{file={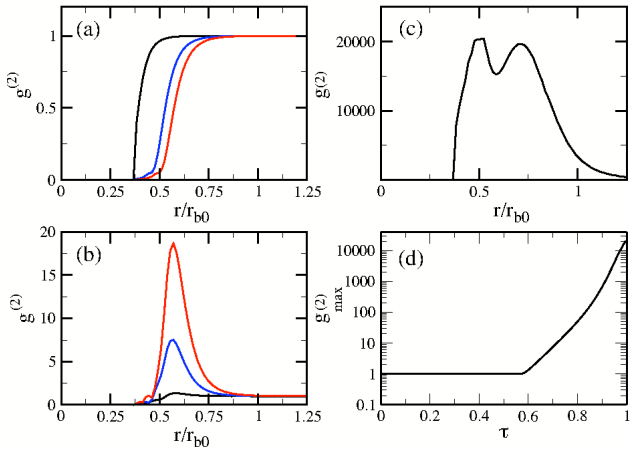},width=0.99\columnwidth} 
\caption{(color online) Spatial atomic correlations during an echo sequence with $\rho_{0} V_{b0}=5.4$. (a) Pair correlation function $g^{(2)}(r)$ during the excitation phase at $\tau=0.05$ (black), $\tau=0.37$ (blue) and $\tau=0.49$ (red). Below the cutoff length $\sub{r}{cut}$, $g^{(2)}$ is set to zero \cite{footnote:numerics2}. (b) The same during the de-excitation phase at $\tau=0.6$ (black), $\tau=0.7$ (blue) and $\tau=0.75$ (red). (c) Final shape of the pair correlation function after the pulse ($\tau=1$). (d) Spatial maximum of pair correlations. The location of the time-samples in (a-c) relative to the pulse can be seen clearly in \fref{Echo_numbers}. 
\label{Echo_tevol}}
\end{figure}

In an echo-sequence, we first excite Rydberg states with Rabi-frequency $\Omega$ for some duration $\tau/2$, followed by a $\pi$-phase shift of the laser and hence Rabi coupling $-\Omega$ for further time $\tau/2$. These parameters are defined in \sref{Omegaexp}. For the full quantum dynamics as in \sref{ExcDyn}, we randomly distribute $N_{0}$ atoms, solve the SE and obtain one correlation function defined by \eref{g2defn} for each pair of atoms. These are binned according to the separation $r$ of those atoms. To minimize finite size effects, we only consider atoms (pairs) inside a central cube of side length $L/2$ within our simulation volume to calculate $N_{e}$ ($g^{(2)}$).

In order to obtain a spatial correlation function, we first calculate $g^{(2)}(i,j)$ according to \eref{g2defn} for each of these atom pairs. We then determine
\begin{align}
g^{(2)}(r)=\bar{\sum}_{i,j}g^{(2)}_{i,j}/N(r),
\label{g2ofr}
\end{align}
where $\bar{\sum}_{i,j}$ denotes a double sum over all atoms that fulfill $|x_{i}-x_{j}|\in [r, r +\Delta r)$ for a bin size $\Delta r$, and $N(r)$ is the number of atom pairs that fall into each respective bin.

Thus averaging over the whole sample and further over a large number of realizations of the spatial atomic distribution \cite{footnote:trajectories}, we obtain a spatial correlation function $g^{(2)}(r)$. Our results from numerical solutions \cite{footnote:numerics2} of \eref{SE1} are shown in \fref{Echo_numbers} and \fref{Echo_tevol}. 

Soon after initiating a transfer of atoms to the Rydberg state, the probability to find a pair of excited atoms closer than the sharply defined instantaneous blockade radius $r_{b}(t)$ is essentially zero. This is due to strong van-der-Waals interactions of those pairs, shifting doubly excited states out of resonance. The radius $r_{b}(t)$ grows in time, since the longer pulse duration allows an increasingly finer energy resolution. The Rydberg atom number in the echo-sequence does not reach the saturation limit, hence $r_{b}(t)$ does not reach its equilibrium extension ($r_{b0}$) but continues to grow throughout the excitation period [\fref{Echo_tevol}, panel ($a$)]. Following the flip of the phase of the Rabi-coupling at $\tau=0.5$, $g^{(2)}$ develops a dominant peak just outside the blockade radius [panel ($b$)]. The height of this peak increases as more and more atoms are transferred back to the ground state [panel ($d$)]. Just before the end of the pulse, for densities as low as in \fref{Echo_tevol}, a dip appears in the correlation peak like that shown in panel (c). This feature starts to occur at the time when the probability for a single excitation in the system drops below that for double excitation.

The underlying physics requires only binary interaction: Consider a pair of atoms separated by some distance $r$. Let us write the quantum state of this pair as 
\begin{align}
|\Psi \rangle & = c_{gg} |g_{1} g_{2} \rangle + c_{eg} |e_{1} g_{2} \rangle + c_{ge} |g_{1} e_{2} \rangle + c_{ee} |e_{1} e_{2} \rangle.
\label{simplestate}
\end{align}
Only in the state $|e_{1} e_{2} \rangle$ where both atoms are excited to a Rydberg level do they experience any interaction. The correlation function in state \bref{simplestate} is
\begin{align}
 g^{(2)}(1,2) &\equiv \frac{\langle \sigma_{ee}^{1}(\tau)\sigma_{ee}^{2} (\tau)\rangle}{\langle\sigma_{ee}^{1} (\tau)\rangle\langle\sigma_{ee}^{2}(\tau) \rangle}
\CR 
%=|c_{ee}|^2/[(|c_{eg}|^2 + |c_{ee}|^2)(|c_{ge}|^2 + |c_{ee}|^2)].
 &=\frac{|c_{ee}|^2}{(|c_{eg}|^2 + |c_{ee}|^2)(|c_{ge}|^2 + |c_{ee}|^2)}.
\label{g2_binarysystem}
\end{align}
Now consider $g^{(2)}$ after the pulse for three different atomic separations: (i) If the atoms are very close $r \lesssim \sub{r}{b}(t)$ double excitation can be considered fully suppressed. Hence $|c_{ee}|= 0$ and $g^{(2)}=0$. (ii) If the atoms are very far apart the interaction can have no effect. We know then $g^{(2)}=1$. From \eref{g2_binarysystem} this can be understood since without interactions and for small excited fractions $|c_{ee}| \approx |c_{eg}|^2= |c_{ge}|^2$ and $|c_{eg}|^2 \gg |c_{ee}|^2$. (iii) In the intermediate range we can neither neglect double excitations nor interaction. Consider the very end of the pulse: The amplitudes $c_{eg}$ and $c_{ge}$ have returned to their initial value of zero after the echo-pulse. In contrast $|c_{ee}|$ is non-zero due to the dephasing. We then see that the atomic correlation function scales as $g^{(2)}=1/|c_{ee}|^2$, which is larger than one. 

As shown in \fref{Echo_tevol}, the distances $r$ with pairing correlations during the second half of the pulse are those where correlations change, as $r$ increases, from blockaded ($g^{(2)}=0$) to uncorrelated ($g^{(2)}=1$) during the first half of the pulse. We will call the spherical shell around each atom where neighbors have these distances the \emph{partial blockade shell}. 

We find that the position of the maximum of the correlation function during the second half of the pulse depends only weakly on time.

%%%%%%%%%%%%%%%%%%%%%%%%%%%%%%%%%%%
\section{Signatures of pairing}
\label{signatures}

In this section we quantify to what extent the correlation dynamics presented in the previous section allows control over the nearest neighbor distribution in a Rydberg gas. We further discuss observables that are easier accessible experimentally than the density-density correlations themselves. 

We consider the fraction $R$ of excited atoms separated from their nearest neighbor by a distance in the interval $[r_{0,}r_{0}+d]$, chosen to contain the peaks in \frefp{Echo_tevol}{a}. This is given by 
\begin{align}
R&= \frac{\sum_{\bv n \neq \bv 0}|c_{\bv n}|^2  f(\bv n)}{\sum_{\bv n \neq \bv 0}|c_{\bv n}|^2},
\label{Rfrac_direct}
\end{align}
where $f(\bv n)$ is the fraction of excited atoms in the many-body basis state $|\bv n \rangle $ with at least one excited neighbor in the chosen interval.

\begin{figure}[htb]
\centering
\epsfig{file={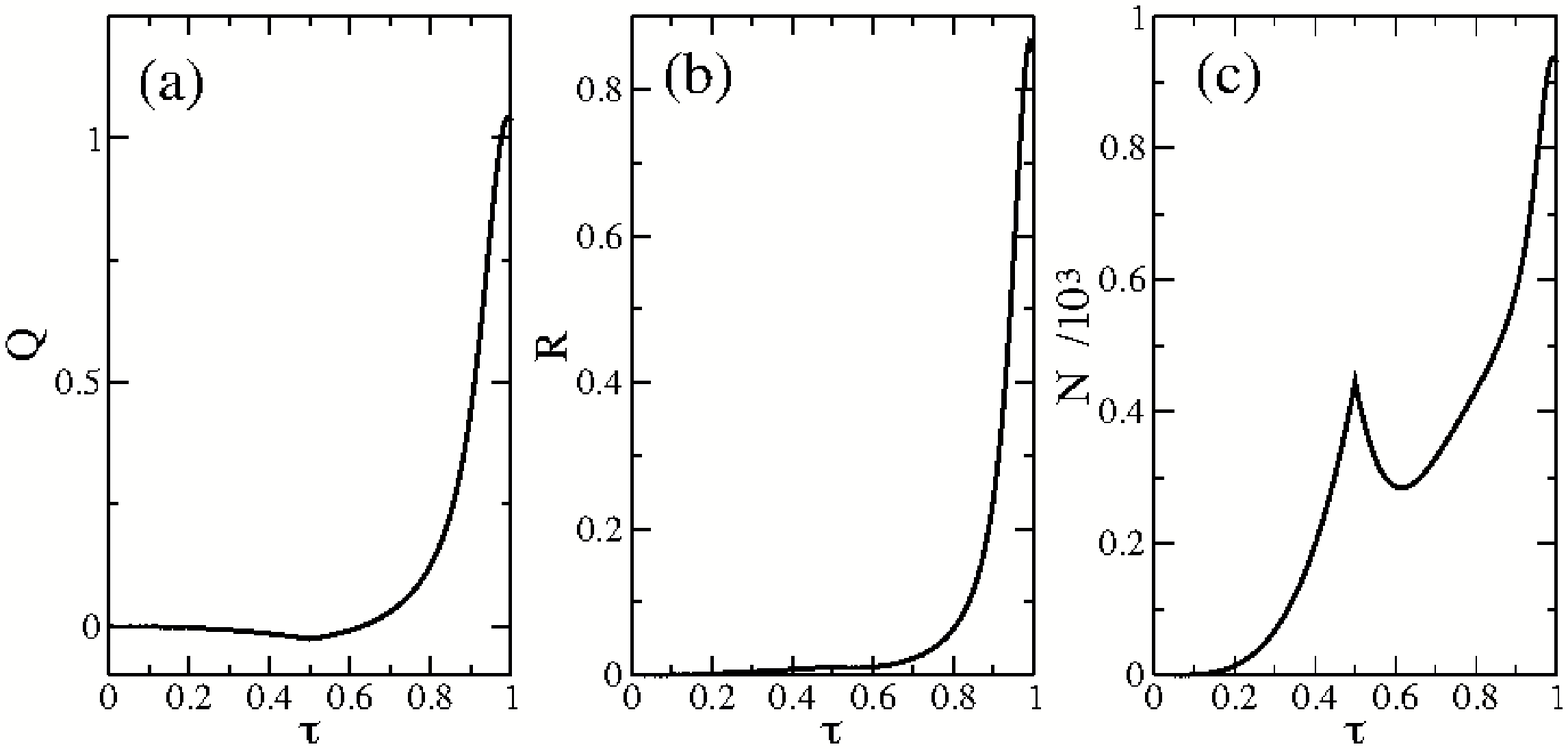},width=0.99\columnwidth}  
\caption{Visibility of the pairing effect for a density with $\rho V_{b0}=26.2$. (a) Mandel-Q parameter.  (b) Fraction $R$ of remnant atoms within the preferred distance peak, using \eref{Rfrac_direct}. (c) Total number of paired, excited atoms, extrapolated \cite{footnote:extrapolation} as if there was a total number of $N_{g}=1 \times 10^7$ atoms initially.
\label{Echo_Mandel}}
\end{figure}
%
%$R$ can also be approximated by \cite{torquato:correls}
%%
%\begin{align}
%R&= 4\pi \int_{r_{0}}^{r_{0}+d}\rho_e(r) r^2 g^{(2)}(r).
%%R&= 4\pi \int_{r_{0}}^{r_{0}+d}\!\!\! \rho_e(r) r^2 g^2(r) \exp[-4\pi \int_{0}^r  \!\!\!   \rho_e(r) r^2 g^2(r) ]  dr 
%% \CR
%%&\approx 4\pi \int_{r_{0}}^{r_{0}+d}\rho_e(r) r^2 g^2(r) .
%\label{Rfrac}
%\end{align}
%%
%Only two-particle correlations are considered in \eref{Rfrac} and hence it is only applicable as long as $R$ is well below $1$. Otherwise the blockade between several atoms in the partial blockade shell (over which one integrates) affects the result. 

For a situation as in \fref{Echo_tevol} we choose $r_{0}=0.5 r_{b0}$ and $d=0.25 r_{b0}$. We show in \frefp{Echo_Mandel}{b} that almost all atoms are paired up before the end of the pulse. In the initial phase, $R$ remains nearly zero owing to the predominance of single excited atoms without neighbor. In panel $c$ we show the total number of paired, excited atoms, obtained by multiplying the paired fraction $R(t)$ with the Rydberg-number $N_{e}(t)$. We then rescale the obtained number corresponding to a total initial number of $10^7$ atoms for illustrative purposes. It may seem counter-intuitive that the number of paired atoms rises even during the second half of the pulse when atoms are predominantly deexcited. Note however, that whether or not the Rabi-coupling causes excitations or de-excitations depends on the relative populations \emph{and} phases. The continuous increase of $R N_{e}(t)$ is again due to interaction induced decoherence.

Finally we describe two possibilities to experimentally detect the correlation dynamics: (i) Via the effect of correlations on the number uncertainty in the Rydberg fraction \cite{cenap:mandelQ} and (ii) via field ionization of the paired Rydberg atoms.

(i) Strong deviations from an uncorrelated state have been shown to affect the number statistics of the excited state fraction  \cite{cenap:mandelQ, cubel:statistics}. This is well captured in the Mandel-Q parameter \cite{mandel:Q}
\begin{align}
Q&= \frac{\expec{\hat{N}_{e}^2}-\expec{\hat{N}_{e}}^2}{\expec{\hat{N}_{e}}}-1= \int  d^3r \rho_{e}(r)[ g^{(2)}(r)-1],
\label{mandelQ}
\end{align}
which can be experimentally determined from the Rydberg atom counting statistics. We show the Q-factor in \frefp{Echo_Mandel}{a}. In comparison with the evolution of $R$, we see that the Q-factor becomes large when most Rydberg atoms are paired up. $Q\sim 1$ already represents a sizable deviation of the atom statistics from a Poissonian distribution. 

(ii) The Q-factor captures only integrated properties of the correlation function. To obtain information about the spatial shape of atomic correlations after an echo pulse, the paired Rydberg atoms could be field ionized. The potential energy of the ions due to Coulomb repulsion $\sub{E}{coul}\approx e^2/r_{0}4\pi\epsilon_{0}$ would subsequently be converted into kinetic energy. Since only a small number of atoms occupies Rydberg states, one could hope that the Rydberg fraction is sufficiently dilute for Coulomb-scattering to leave the initial kinetic energy distribution essentially unchanged. The peak in the nearest-neighbor distribution function $\sim r^2 g^{(2)}(r)$ then translates into a well visible maximum of the measured kinetic energy spectrum of the ions. For $\rho V_{b0}=26.2$ the energy at the maximum is $0.5$ meV.

%%%%%%%%%%%%%%%%%%%%%%%%%%%%%%%%%%%
\section{Varying the spatial atomic distribution}
\label{interact_inhomog}

\begin{figure}
\centering
\epsfig{file={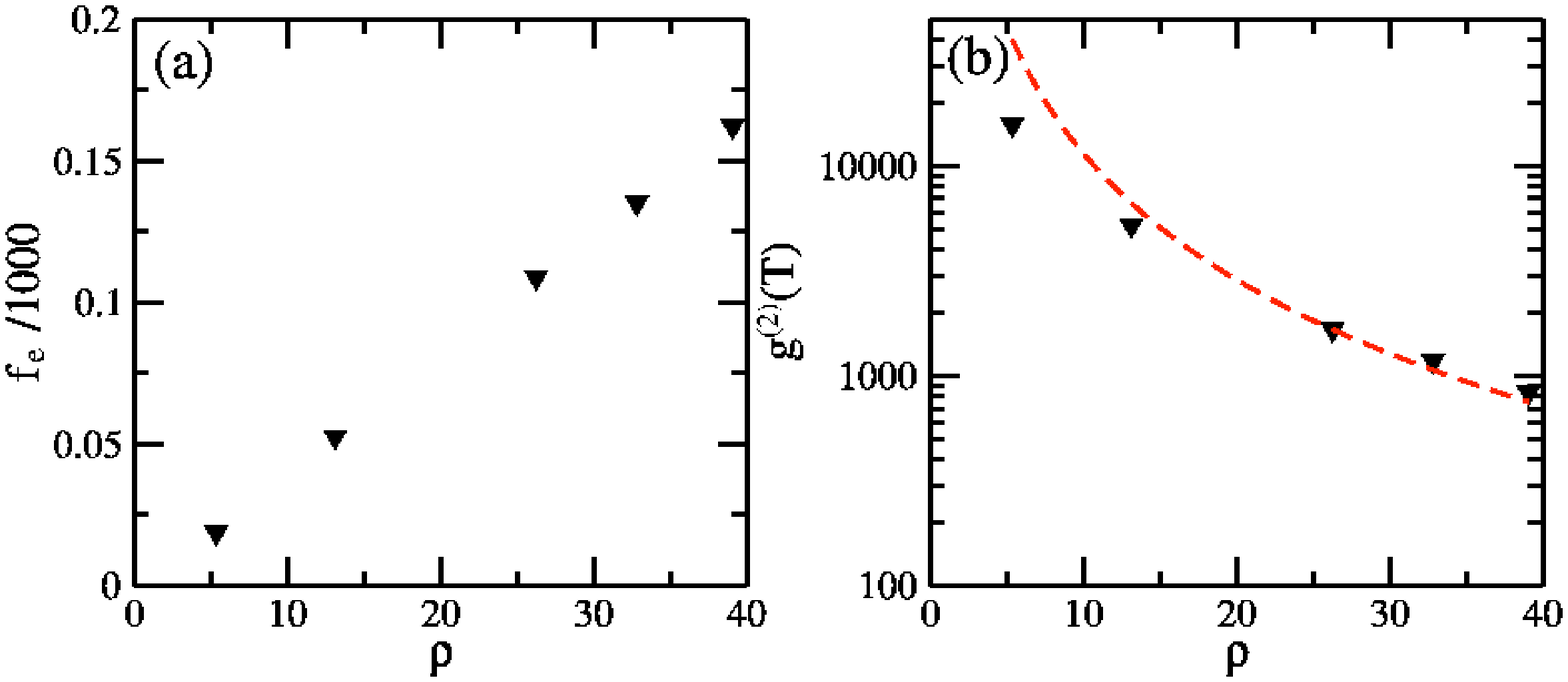},width=0.99\columnwidth} 
\caption{(color online) Dependence of echo signal and final correlation peak on atomic density. (a) Final Rydberg fraction at $t=T$. These datapoints are very well described by \eref{echosig}. (b) ($\blacktriangledown$) Peak height of $g^{(2)}$ at $t=T$. The (red-dashed) curved shows the functional form $g^{(2)}(\rho)=[\rho_{0}/\rho]^2 g^{(2)}(\rho_{0})$ with reference density $\rho_{0}=26.2$. This is motivated by a simple model explained in the text.
\label{Echo_density}}
\end{figure}

The scenarios shown in the previous section are relatively weakly blockaded and assume a homogeneous distribution of atoms. In this section we study how an in-homogeneous distribution and changes of the overall density affect our results.
\begin{figure*}
\centering
\epsfig{file={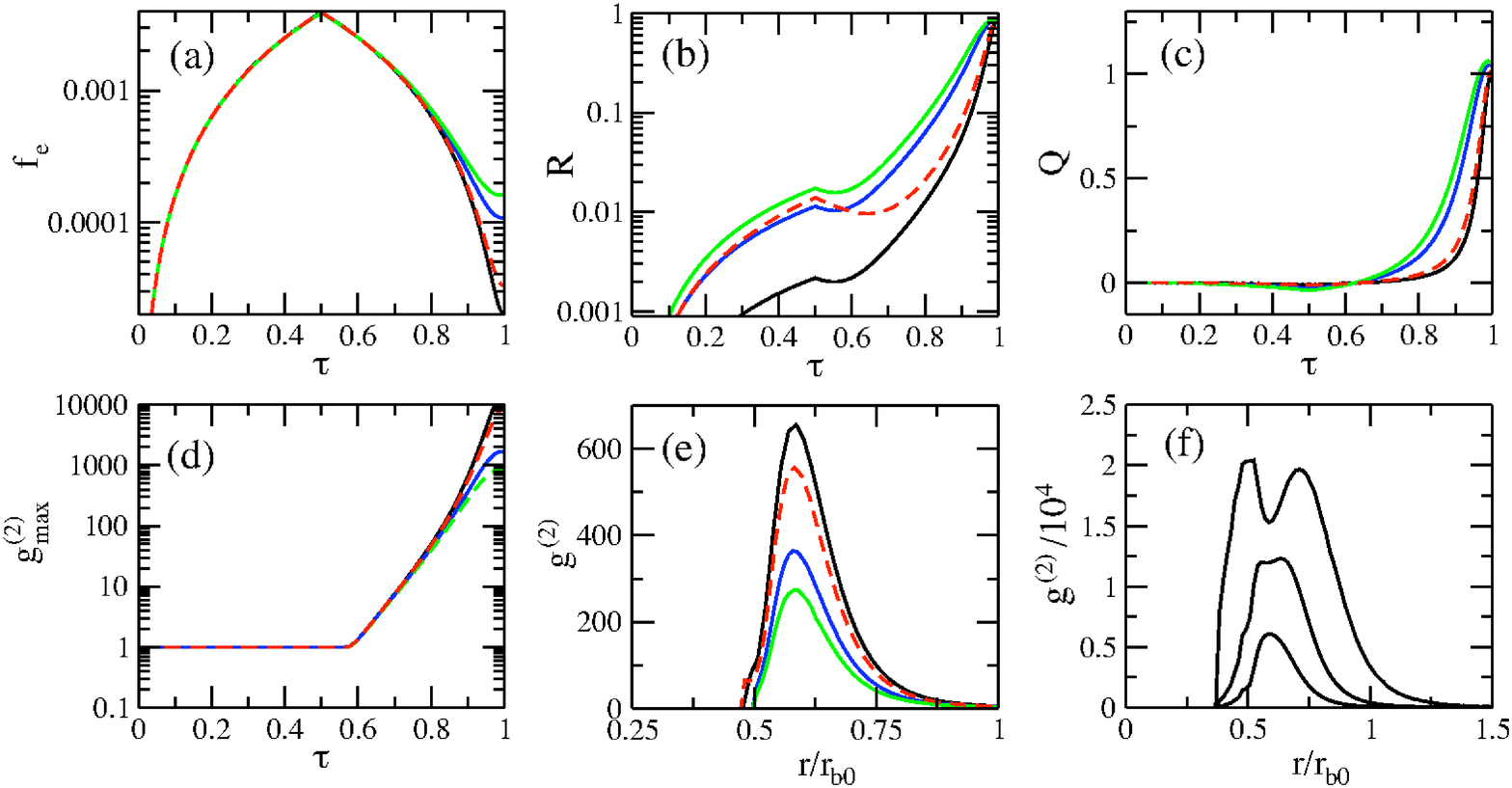},width=1.79\columnwidth} 
\caption{(color online) Effects of inhomogeneities and changes of density on the pairing peak. We show densities $\rho_{0} V_{b0}=5.4$, $26.2$, $39$ in (black, blue, green) respectively, and an inhomogeneous case $\rho(x)=\rho_{0}\exp{[-2r^2/\sigma^2]}$ with $\rho_{0}V_{b0}=26.2$ and $\sigma/r_{b0}=1.6$ (red-dashed). (a) Rydberg fraction $f_{e}$, for increasing homogeneous density the final remnant $f_{e}(\tau=1)$ gets larger. (b) Paired excited fraction according to \eref{Rfrac_direct}. (c) Mandel-Q parameter, see \eref{mandelQ}. (d) Spatial maximum of pair correlations. (e) $g^{(2)}(r)$ at $\tau=0.75$, for increasing homogeneous density, the peak height decreases. (f) Spatial atomic correlations near the end of an echo-pulse in a low density ($\rho_{0} V_{b0}=5.4$) gas. Shown are snapshots at $\tau=0.96$, $\tau=0.98$, and $\tau=1.0$ in order of increasing peak-height.
\label{Echo_interact}}
\end{figure*}
The basic picture is shown in \fref{Echo_interact} and \fref{Echo_density}. In \fref{Echo_interact} we compare a variety of different densities. Prominent features of this data are compared more directly in \fref{Echo_density}.

Is is known that the echo-signal, i.e.~the fraction of Rydberg excitations remaining after the pulse, increases as the density is increased \cite{raitzsch:echo}. Using the $\omega$-expansion, we can supply the analytical expression \eref{echosig} for this increase. For the low densities where it is valid, it performs well as can be seen in \frefp{Echo_density}{a}. Other effects of increasing density are a decrease of the maximum correlations and an increase of the Mandel-Q factor for intermediate times ($\tau\sim3/4$). Despite the decrease of the correlation peak height, we find that the paired fraction $R$, defined by \eref{Rfrac_direct} almost reaches unity after the pulse, regardless of the density. This is shown in \frefp{Echo_interact}{e}. We also show the development of the dip in the correlation peak in \frefp{Echo_interact}{f}. We only see this feature for the lowest densities considered.

The reduction of correlation strength with increasing density can be understood from the nature of the distribution of excitations near the end of the pulse, using only classical arguments. Let us assume a homogenous distribution of pairs of excited atoms within a volume $V$, with a fixed distance $r_{0}$ between the partners of each pair. The distance $r_{0}$ then corresponds to the location of our pairing peak, ignoring its finite width. The orientation of pairs in space shall be isotropic. Overall we thus have a distribution of positions for each pair:
\begin{align}
f(\bv{R},\bv{r})=\frac{1}{4 \pi V}\delta(|\bv{r}|-r_{0}).
\label{pairdens}
\end{align}
Here, $\bv{R}$ denotes the centre-of-mass position of a pair and $\bv{r}$ its relative co-ordinate. Assuming a total number of $M$ pairs we have a pair density $n_{p}=M/V$ and an excited atom density $n_{e}=2 M/V$. The positions of the pairs themselves are correlated due to the dipole blockade, we thus assume that the centers of the pairs must be separated by more than a certain radius $r_{b}$.
The classical correlation function corresponding to our $g^{(2)}$ is
\begin{align}
&\bar{g}^{(2)}(r) = 
\label{classicalg2}
\nnl
&\int d\Omega
\frac{
E(\bv{r_1}=\bv{x_0};\bv{r_2}=\bv{x_0}+\bv{y}\mbox{, }|\bv{y}|=r)
}{
E(\bv{r_1}=\bv{x_0})E(\bv{r_2}=\bv{x_0}+\bv{y}\mbox{, }|\bv{y}|=r)
}.
\end{align}
In the numerator we have the expectation value for the number of joint occupations of the locations $r_{1}$ and $r_{2}$ by excited atoms. In the denominator $E(\bv{r_1}=\bv{x_0})$ denotes the expectation value for the number of excited atoms at location $x_{0}$, which is $2 M/V$. The integration $d\Omega$ is over the solid angle of orientations of $\bv{y}$ with respect to $\bv{x_0}$. 
We are now interested exclusively in the value of the correlation function at the pairing peak, located at $r=r_{0}$. For simplicity we assume contributions to the numerator of \eref{classicalg2} stem only from cases where the atoms at both locations were constituents of the same pair. This should be justified when the pairs are sufficiently dilute that their mean distance is much larger than $r_{0}$. We can then write $E(\bv{r_1}=\bv{x_0};\bv{r_2}=\bv{x_0}+\bv{y}\mbox{, }|\bv{y}|=r)=M f{((\bv{r_{1}}+\bv{r_{2}})/2,\bv{r_{1}-\bv{r_{2}}}})$.

With these simplifying assumptions, we find \eref{classicalg2} equals $V/4 M\sim 1/n_{p}$ for the simple pair distribution described above. Using $n_{p}\sim n_{e}$, $n_{e}=f_{e}\rho$ and $f_{e}\sim \rho$ from the $\omega$-expansion, we find overall that the correlation strength scales like $\rho^{-2}$. This behavior is roughly confirmed by the simulation results as shown in \frefp{Echo_density}{b}. It is consistent with a paired fraction $R$ that shows almost no density dependence near the end of the pulse: For each excitation we calculate $N=\int_{r_0}^{r_0+ d} r^2 g^{(2)}(r) n_{e}(r) dr$
to obtain the number of excited neighbors in an interval $[r_{0}, r_{0}+ d]$. Since $n_{e}$ scales like $\rho^2$, while $g^{(2)}$ scales like $\rho^{-2}$, we can understand how the final paired fraction can remain almost the same even though the density is varied.

We verified that the relation $g^{(2)}(r_{0},\tau)\sim \rho^{-2}$ remains qualitatively unchanged if we model similar echo-pulses in a fictitious system with a $1/r^4$ long range interaction. This lends further support to the simple explanation in terms of the density of available pairs.

Now consider a case with $N_{0}=50$ atoms inhomogeneously distributed with atomic density $\rho(x)= \rho_0\exp{[-2r^2/\sigma^2]}$ [case {\cased} in \tref{scenarios}]. The peak-density $\rho_0$ is chosen as for case {\caseb}.  \fref{Echo_interact} includes the correlation function averaged over all atomic pairs in the cloud for this case. It is determined in the same manner as described in \sref{correldyn}.
 One could expect that the inhomogeneity washes out the signal in the correlation function. Instead its visibility is even better than for the homogenous case with equal peak-density, owing to the presence of low density regions in the atomic cloud. For lower density the pairing effect is more prominent as we already argued.
We also do not expect a strong spatial variation of the preferred distance throughout the cloud, since the blockade radius $r_{b}(t)$ depends only weakly on the atomic density (i.e.~the saturated many-body blockade radius scales like $\bar{r}_{b}\sim\rho_{g}^{2/5} \Omega$ \cite{heidemann:rydberg}). Hence the location of the correlation peak, situated near $r_{b}(\tau/2)$, is expected to vary only little throughout the sample. Consequently there are only small effects of spatial averaging on the final result in the inhomogenous case considered here. 
To fully exclude that density inhomogeneities in an experiment would suppress the signature reported here, one ideally should consider a sample whose width greatly exceeds the characteristic range where the correlation peak is formed ($\sigma \gg r_{0}$). This is however computationally intractable.

%%%%%%%%%%%%%%%%%%%%%%%%%%%%%%%%%%%
\section{Comparison with $\omega$-expansion}
\label{omegaexp}

The simple expression \eref{g2anal} for two-body correlations in the $\omega$-expansion does not depend on density and hence cannot capture its effect on correlation dynamics as shown in \fref{Echo_interact}. However, for small densities it compares quite well with the substantially more involved exact Schr{\"o}dinger evolution. This can be expected from \eref{peexp}: For $N_{b}=40$, the series is well described by its first two terms until $\Omega t=0.25$.

\begin{figure}[htb]
\centering
\epsfig{file={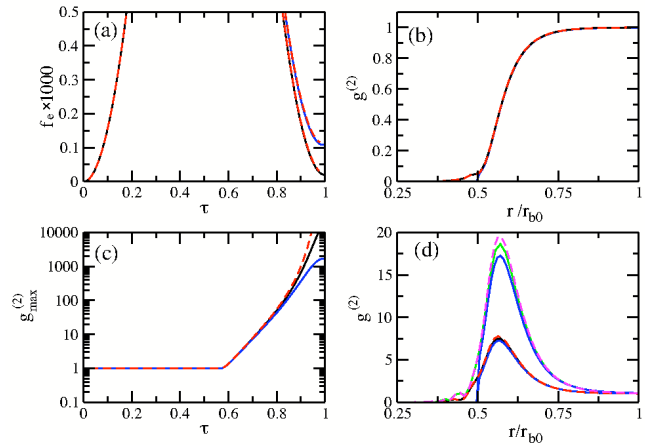},width=0.99\columnwidth}  
\caption{(color online) Comparison of the $\omega$-expansion (red-dashed) with exact solutions of the SE  for $\rho_{0} V_{b0}=5.4$ (solid black)  and $\rho_{0} V_{b0}=26.2$ (solid blue). (a) Excited state fractions. (b) Pair correlation function $g^{(2)}(r)$ at $\tau=0.49$. (d)  Pair correlation function $g^{(2)}(r)$ at $\tau=0.6$ and $\tau=0.75$. The later time has a higher pairing peak. For $\tau=0.6$ we used the same color scheme as in the other panels, for $\tau=0.75$ the $\omega$-expansion result is (magenta-dashed) and the
exact solution is (solid green). (c) Spatial maximum of pair correlations. 
\label{omega_comparison}}
\end{figure}

We can see in \frefp{omega_comparison}{a} and \frefp{Echo_density}{a} that the NLO result for the Rydberg fraction [\eref{PeNLO}] gives good quantitative results for the cases considered here. 
For correlations, we only have the LO expression \eref{g2anal}. We see that this approximation describes correlations in the low density case $\rho_{0} V_{b0}=5.4$ quite well, while quantitative deviations appear, once the density becomes as high as in the case with $\rho_{0} V_{b0}= 26.2$. Near the very end of the pulse also qualitative differences in the shape of the correlation function are present, since \eref{g2anal} cannot describe the dip seen in \frefp{Echo_interact}{f}.  

We note that in particular for the cases that show the most dramatic correlation dynamics through an echo pulse, those with a low density, the $\omega$-expansion provides useful results.

%%%%%%%%%%%%%%%%%%%%%%%%%%%%%%%%%%%
\section{Conclusions}
\label{conclusions}

We have shown that an echo-type excitation sequence as employed in the experiments \cite{raitzsch:echo,younge:echo} can be used to create Rydberg gases with non-standard nearest neighbor-distribution functions. After the pulse the vast majority of excited atoms possesses a neighbor in a fairly narrow interval around some distance $r_{0}$. 
Variations of the interaction-strength and pulse-length can control $r_{0}$. The strength of the correlation signal is proportional to $\rho^{-2}$, where $\rho$ is the atomic density. 
This can be understood in terms of the quantum state after the echo-pulse, independent of the precise form of the interaction.

The described pairing effect in the density-density correlation function is most pronounced for low densities and not too strongly blockaded gases.
However the overall fraction of Rydberg atoms that have a neighbor near the distance $r_{0}$ approaches unity regardless of density. Our results were obtained by direct simulation of ensembles of about a hundred atoms. Further we used these simulations to estimate the range of validity of the first order approximation of the correlation dynamics obtained using the $\omega$-expansion. We find that for weakly blockaded gases it provides useful estimates.

%%%%%%%%%%%%%%%%%%%%%%%%%%%%%%%%
\acknowledgments
PD was supported by the European Community under the contract MEIF-CT-2006-041390.


\begin{thebibliography}{24}
\bibitem[{foo({\natexlab{a}})}]{footnote:numerics2}
\bibinfo{note}{Our simulations use a 8th-9th order adaptive stepsize
  Runge-Kutta Fehlberg method. The cutoff energy $\sub{E}{cut}$ was typically
  taken as the van-der-Waals energy of a pair of atoms at a distance
  $\sub{r}{cut}=0.5 r_{b0}$, $\sub{N}{max}$ was about $4-6$. For case {\casea}
  we used $\sub{r}{cut}=0.38 r_{b0}$. None of the results presented changed
  upon moderate variation of $\sub{E}{cut}$, $\sub{N}{max}$, the time-step
  tolerance or the cubic box length $L$.}
\bibitem[{foo({\natexlab{c}})}]{footnote:extrapolation}
\bibinfo{note}{To extrapolate, we simply scale the total atom number from $84$
  to $10^7$. This assumes that the simulated cube gives a representative sample
  of the dynamics of a large homogenous atomic gas.}
\end{thebibliography}
\end{document}